\newif\iffigs\figstrue

\documentstyle[12pt]{article}

\iffigs
  \input epsf
\else
\fi

\batchmode
  \newfont{\footscrfont}{rsfs10}
  \newfont{\footbbbfont}{msbm10}
\errorstopmode

\newif\ifscrf\scrftrue
\ifx\footscrfont\nullfont
  \scrffalse
\fi

\newif\ifamsf\amsftrue
\ifx\footbbbfont\nullfont
  \amsffalse
\fi

\def\ppnumber{\vbox{\baselineskip16pt\hbox{DUK-TH-94-68}
  \hbox{IASSNS-HEP-94/23}}}
\def\ppdate{April, 1994}
\def\pplogo{\vbox{\kern-\headheight\kern -17pt
\halign{##&##\hfil\cr&{
\ppnumber}\cr\rule{0pt}{2.5ex}&\ppdate\cr}
}}

\makeatletter
\date{}
\def\dedicatory#1{\def\@date{\normalsize\it#1}}
\def\subjclass#1{\def\@thefnmark{}\@footnotetext{1991
    {\it Mathematics Subject Classification.} #1}}
\def\keywords#1{\def\@thefnmark{}\@footnotetext{
    {\it Key words and phrases.} #1}}

\def\ps@firstpage{\ps@empty \def\@oddhead{\hss\pplogo}%
  \let\@evenhead\@oddhead 
}
\def\maketitle{\par
 \begingroup
 \def\thefootnote{\fnsymbol{footnote}}
 \def\@makefnmark{\hbox
 to 0pt{$^{\@thefnmark}$\hss}}
 \if@twocolumn
 \twocolumn[\@maketitle]
 \else \newpage
 \global\@topnum\z@ \@maketitle \fi\thispagestyle{firstpage}\@thanks
 \endgroup
 \setcounter{footnote}{0}
 \let\maketitle\relax
 \let\@maketitle\relax
 \gdef\@thanks{}\gdef\@author{}\gdef\@title{}\let\thanks\relax}

\def\abstract{\if@twocolumn
\section*{Abstract}
\else \small
\begin{center}
{\bf ABSTRACT}
\end{center}
\quotation
\fi}

\def\thebibliography#1{\section*{References\@mkboth
 {REFERENCES}{REFERENCES}}\small\list
 {\arabic{enumi}.}{\settowidth\labelwidth{[#1]}\leftmargin\labelwidth
 \advance\leftmargin\labelsep
 \usecounter{enumi}}
 \def\newblock{\hskip .11em plus .33em minus .07em}
 \sloppy\clubpenalty4000\widowpenalty4000
 \sfcode`\.=1000\relax}

\newif\iffn\fnfalse

\@ifundefined{reset@font}{\let\reset@font\empty}{} 
\long\def\@footnotetext#1{\insert\footins{\reset@font\footnotesize
    \interlinepenalty\interfootnotelinepenalty
    \splittopskip\footnotesep
    \splitmaxdepth \dp\strutbox \floatingpenalty \@MM
    \hsize\columnwidth \@parboxrestore
   \edef\@currentlabel{\csname p@footnote\endcsname\@thefnmark}\@makefntext
    {\rule{\z@}{\footnotesep}\ignorespaces
      \fntrue#1\fnfalse\strut}}}
\makeatother

\ifamsf
  \newfont{\bigbbbfont}{msbm10 scaled\magstep2}
  \newfont{\bbbfont}{msbm10 scaled\magstep1}  
  \newfont{\smallbbbfont}{msbm8}
  \newfont{\tinybbbfont}{msbm6}
  \newfont{\smallfootbbbfont}{msbm7}
  \newfont{\tinyfootbbbfont}{msbm5}
\fi

\ifscrf
  \newfont{\scrfont}{rsfs10 scaled\magstep1}  
  \newfont{\smallscrfont}{rsfs7}
  \newfont{\tinyscrfont}{rsfs7}
  \newfont{\smallfootscrfont}{rsfs7}
  \newfont{\tinyfootscrfont}{rsfs7}
\fi

\ifamsf
  \newcommand{\Bbb}[1]{\iffn
      \mathchoice{\mbox{\footbbbfont #1}}{\mbox{\footbbbfont #1}}
      {\mbox{\smallfootbbbfont #1}}{\mbox{\tinyfootbbbfont #1}}\else
      \mathchoice{\mbox{\bbbfont #1}}{\mbox{\bbbfont #1}}
      {\mbox{\smallbbbfont #1}}{\mbox{\tinybbbfont #1}}\fi}
\else
  \def\bigbbbfont{\bf}
  \def\Bbb{\bf}
\fi

\ifscrf
  \newcommand{\Scr}[1]{\iffn
    \mathchoice{\mbox{\footscrfont #1}}{\mbox{\footscrfont #1}}
    {\mbox{\smallfootscrfont #1}}{\mbox{\tinyfootscrfont #1}}\else
    \mathchoice{\mbox{\scrfont #1}}{\mbox{\scrfont #1}}
    {\mbox{\smallscrfont #1}}{\mbox{\tinyscrfont #1}}\fi}
\else
  \def\Scr{\cal}
\fi

\def\operatorname#1{\mathop{\rm #1}\nolimits}
\def\C{{\Bbb C}}

\def\P{{\Bbb P}}

\def\R{{\Bbb R}}
\def\Z{{\Bbb Z}}

\def\opeq#1{\advance\lineskip#1 \advance\baselineskip#1
	\advance\lineskiplimit#1}
\def\eqalign#1{\null\,\vcenter{\opeq{2.5\jot}\mathsurround=0pt
	\everycr={}\tabskip=0pt
	\halign{\strut\hfil$\displaystyle{##}$&$\displaystyle{{}##}$\hfil
	\crcr#1\crcr}}\,\null}

\def\sm{$\sigma$-model}
\def\nlsm{non-linear \sm}

\def\CY{Calabi-Yau}
\def\Kf{K\"ahler form}

\def\cM{{\Scr M}}

\def\cZ{{\Scr Z}}

\def\cMc{{\hfuzz=100cm\hbox to 0pt{$\;\overline{\phantom{X}}$}\cM}}

\ifscrf
  \def\Tch#1#2{{\Scr T}^{\;\,#1,#2}}
\else
  \def\Tch#1#2{{\cal T}^{#1,#2}}
\fi

\ifamsf
  \def\ltimes{\mathbin{\mbox{\bbbfont\char"6E}}}
\else
  \def\ltimes{.}
\fi

\begin{document}
\setcounter{page}0
\title{\LARGE String Theory\\
on K3 Surfaces\\[10mm]
}
\author{
Paul S. Aspinwall,\\[0.7cm]
\normalsize School of Natural Sciences,\\
\normalsize Institute for Advanced Study,\\
\normalsize Princeton, NJ  08540\\[10mm]
David R. Morrison \\[0.7cm]
\normalsize  Department of Mathematics, \\
\normalsize  Box 90320, \\
\normalsize  Duke University, \\
\normalsize  Durham, NC 27708-0320}

{\hfuzz=10cm\maketitle}

\def\Large{\large}
\def\LARGE{\large\bf}


\begin{abstract}

The moduli space of $N$=(4,4) string theories with a K3 target space
is determined, establishing in particular that the discrete symmetry
group is the full integral orthogonal group of an even unimodular
lattice of signature (4,20).
The method combines an analysis of the classical theory of K3
moduli spaces with mirror symmetry. A description of the moduli
space is also presented from the viewpoint of quantum geometry,
and consequences are drawn concerning mirror symmetry for algebraic
K3 surfaces.

\end{abstract}

\vfil\break

\section{Introduction}		\label{s:intro}

In recent years, \CY\ manifolds have
received great attention in the string literature. This is mainly because
compactification on
such spaces may be used to reduce the number of dimensions in models built
 from the intrinsically ten dimensional critical superstring \cite{CHSW:}.
The focus has largely been on the case of complex dimension $d=3$,
since the corresponding compactification is to four space-time dimensions.
(It is also the case that, in many respects, within topological
field theory on \CY\ target spaces
the case $d=3$ plays a special r\^ole
\cite{W:aspects,BCOV:big}.)
However,
one might expect interesting properties for other values of $d$ as well.

In this paper we analyze the
case $d=2$ whose special features derive from the fact that the topology
of the target must be either
a torus or a K3 surface. The classification of
toroidal target spaces has been known for some time \cite{N:torus,HNW:torus}.
The local geometry of the moduli space of conformal field
theories with K3 target space topology
was understood shortly afterwards \cite{Sei:K3} and although some
aspects of the global form of the moduli space have been conjectured
and studied \cite{Sei:K3,Vafa:qu,GS:sym}, more precise statements
concerning this question have remained elusive.

We will use mirror symmetry to address this question of
the form of the moduli space. We find that, given a couple of minor
assumptions, combining a \sm\ analysis near the field theory
limit with the study of a mirror symmetry transformation
is sufficient to give the precise form of the moduli
space. This paper is not intended to provide the full mathematical
exposition of the analysis required --- for that the reader is
referred to \cite{AM:K3m}. Here we only give a brief summary of the
methods used and conclusions reached by the analysis.

The analysis of the moduli space of string theories on a K3 surface
differs markedly from that of
the more familiar \CY\ moduli spaces related to string compactifications
to four dimensions.
Firstly, in studying the case where the target space $X$ is a \CY\ threefold,
the assumption $h^{2,0}=0$ is usually made.
Under this assumption,  the
deformations of complex structure of $X$ and of the (complexified)
\Kf\ ``decouple'' to give the complete moduli space a local product structure
\cite{Cand:mir} (at least over generic points in the moduli space).
These two types of deformations may then be studied independently.
For a K3 surface we have $h^{2,0}=1$ and the above structure is lost.
We are thus forced to analyze deformations of complex structure and
deformations of \Kf\ together.

In another respect however the K3 case
is simpler --- the $N$=(2,2) superconformal invariance of a string
with \CY\ target space is extended to $N$=(4,4)
supersymmetry in the case $d=2$
\cite{Eg:K3}.
This is equivalent to the geometric statement that a \CY\ manifold in
complex dimension two admits a hyperk\"ahler metric. This $N$=(4,4)
structure serves to fix the local form of the moduli space
completely. This can be contrasted with the case $d=3$ where the
extended chiral algebra
\cite{Odake:} has yet to provide much insight into the classification
of string theories.

As in the case $d=3$, the local dimension of the moduli space of
conformal field theories can be matched to the dimension of the moduli
space of the geometrical objects representing the target space if we
regard the ``geometrical object'' as including, in
addition to the target space metric, the specification of
a $B$-field, i.e., an
element of the second real (de Rham) cohomology group of the target. In the
case
$d=3$, the $B$-field naturally combined with the K\"ahler form to
provide a complexified K\"ahler form. In the case of K3 this is
clearly impossible --- $B$ lives in a 22-dimensional space whereas the
dimension of the space of K\"ahler forms varies with the complex
structure but is at most 20. Despite this fact we will still find that
a beautiful structure arises when the $B$-field is included in the
moduli space.

At the topological level, mirror symmetry might appear rather trivial
for a K3 surface since the mirror of a K3 surface (equipped with metric
and $B$-field) is another such K3
surface. However, the geometric data of the
mirror K3 surface thus obtained is not, in
general, isomorphic to the original. Thus the mirror map acts as a
non-trivial automorphism on the ``classical'' moduli space of K3 surfaces
(i.e., the \sm\ moduli space of metrics and $B$-fields).
The mirror map in this context was first studied long ago in
\cite{Mart:CCC} and some of the observations in that paper lead to
some of the methods used here.

Although the purpose of this paper is to concentrate on the case where the
target space is a K3 surface, it should be noted that much of
what follows applies to any target space with $h^{2,0}=1$. The only
steps in the following argument which will not be directly applicable
to the determination of the moduli space of this more general case are the
explicit form of the moduli space of
Einstein metrics, and the particular self-mirror K3 theory used.

In section \ref{s:mod} we will present the outline of the construction
of the moduli space of $N$=(4,4) and $N$=(2,2) theories. In section
\ref{s:coh} we discuss the interpretation of the moduli space in terms
of the space of total cohomology of K3. In section \ref{s:alg},
some  aspects of mirror
symmetry on algebraic K3 surfaces are  discussed.


\section{The Moduli Space}	\label{s:mod}

Let us first fix some notations. $\Gamma\backslash G/H$ denotes the double
coset
space resulting from dividing the group $G$ by the right-action of
$H$ and left-action of $\Gamma$. ($G$ and $H$ will always be continuous Lie
groups and $\Gamma$ will be discrete.) $\Lambda^{a,b}$ denotes the unique
(up to isomorphism) even
self-dual lattice in $(a+b)$-dimensional space with signature $(a,b)$,
when $ab\ne0$.
Basis vectors may be taken such that the inner product has the
following form:
\begin{equation}
  \Lambda^{a,b}:\langle,\rangle\cong
\underbrace{(-E_8)\oplus(-E_8)\oplus\ldots}_{n\rm\;times}
\oplus H\oplus H\oplus\ldots,
\end{equation}
where $b-a=8n$
(one uses $E_8$ in place of $-E_8$ when $b-a<0$),
$E_8$ is the Cartan matrix of the Lie algebra $E_8$,
and $H$ is the hyperbolic plane:
\begin{equation}
 H = \left(\begin{array}{cc}0&1\\1&0\end{array}\right). \label{eq:H}
\end{equation}
We will use $X$ to denote a specific smooth K3 surface. One can show that
the intersection pairing on
$H^2(X,\Z)$
gives it the structure of a lattice isomorphic to
$\Lambda^{3,19}$ (see for example \cite{BPV:}).
Let $\R^{a,b}$ be a $(a+b)$-dimensional space with an inner product
of signature $(a,b)$.
$O(a,b)$ is the orthogonal group on $\R^{a,b}$ and $O(\Lambda^{a,b})$
is the subgroup preserving $\Lambda^{a,b}\subset \R^{a,b}$. Let
\begin{equation}
 \Tch ab=O(a,b)/\left(O(a)\times O(b)\right).
\end{equation}
We can identify $\Tch ab$ as the set of space-like $a$-planes in
$\R^{a,b}$ (i.e., $a$-planes on which the inner product is
positive-definite).  $\Tch ab$ can also be regarded as one of the
two connected components of the set of {\em oriented}\/ space-like
$a$-planes.  Under this latter interpretation, we can write
\begin{equation}
\Tch ab=O^+(a,b)/\left(SO(a)\times O(b)\right),
\end{equation}
where $O^+(a,b)$ is the index 2 subgroup of $O(a,b)$ with the
same connected components as $SO(a)\times O(b)$.

It was shown in \cite{Sei:K3} using arguments from supergravity
\cite{deRoo:} that the required moduli space, $\cM$, of conformal field
theories on a K3 surface is locally of the
form $\Tch 4{20}$. This is probably best understood  from
the argument
presented in \cite{Cec:nonp}, which we now review.
The $N$=(4,4) superconformal algebra
contains affine $SU(2)$ algebras in both the left and right sectors.
This symmetry acts on the marginal operators spanning the tangent
spaces of $\cM$. The existence of such a symmetry
 restricts the form of the holonomy group of the
Zamolodchikov metric of $\cM$.
The restriction is so severe, in fact, that
it then follows from Berger's classification \cite{Berg:}
of holonomy groups that either $\cM$ is locally isomorphic to a
quaternionic symmetric space, or that the holonomy is reducible.
Moreover, the
non-flatness of the Zamolodchikov metric is enough to rule out the
reducible holonomy case.
A simple analysis of any conformal field theory giving rise
to a K3 target space tells us that $\cM$ has  real dimension 80
\cite{Sei:K3,Eg:K3}.\footnote{One way to do this count is as
follows:  the space of complex structures has complex dimension 20
(real dimension 40), the space of K\"ahler forms has real dimension
20, and the space of $B$-fields has real dimension 22.  However, for each
Ricci-flat metric on the K3 surface there is an $S^2$ of complex
structures, so the overall dimension count is $40+20+22-2=80$.}
This,
together with the known classification of quaternionic symmetric
spaces (cf.~\cite{Wolf:}), completes the proof.

Let us now make two assumptions about the form of $\cM$, both of which
we consider to be quite reasonable. Firstly we assume $\cM$ is
geodesically complete\footnote{As we will point out below, we must
include orbifold points in our moduli space in order to ensure this
geodesic completeness.}
 and secondly we assume $\cM$ to be Hausdorff.
While some non-Hausdorff moduli spaces have appeared in string theory
\cite{Moore:nH}, this happened in the context of a target space of
indefinite signature. There is no reason to believe that something as
unpleasant as this should happen in the case of a K3 surface.
It then follows that
\begin{equation}
  \cM\cong \Gamma\backslash \Tch 4{20},
\end{equation}
for some group $\Gamma$ acting discretely on $\Tch 4{20}$.
Thus to complete the description of $\cM$, we only need to find $\Gamma$.

We begin our determination of $\Gamma$ by analyzing the
classical form of the moduli
space of K3 surfaces. Using the techniques of \cite{Borel-Serre}, one may
decompose $\Tch 4{20}$ as
\begin{equation}
  \Tch4{20} \cong \R_+\times\Tch3{19}\times \R^{3,19},
		\label{eq:decomp}
\end{equation}
where $\R_+$ is the half-line of positive real numbers which we
parameterize by $\lambda$.
The decomposition depends on a choice of null vector $v\in\R^{4,20}$,
and the space $v^\perp/\R v$ provides the $\R^{3,19}$ on which
$O(3,19)$ acts and on which $\Tch 3{19}$ is based.
The space $\Tch 3{19}$ is known to be isomorphic to
the Teichm\"uller space for Einstein metrics of volume one on a K3 surface
(and their orbifold limits)
\cite{Mor:Katata,Kobayashi-Todorov,Anderson}. It is natural then to
identify $\lambda$ as giving the size of the K3 surface and $\R^{3,19}$
as the moduli space of $B$-fields on the target space. These
identifications may be established by looking at the metrics on the
above mentioned spaces. The Zamolodchikov metric on $\Tch 4{20}$ is
known to be the left-invariant metric. From \cite{Borel:stable} this induces
a metric on each of the terms on the right hand of (\ref{eq:decomp})
where this should now be viewed as a ``warped product'', i.e., the
metric does not respect the product structure. One may now show
(using \cite{Anderson}) that
the metrics induced are those given precisely by the extended
Weyl-Petersson metric in the sense of \cite{CGH:WP}. Such an isometric
identification allows us to
identify every point in $\Tch4{20}$ with an Einstein metric on a K3
surface (or orbifold metric)
together with a $B$-field. This can be taken as another
version of the statement that the \nlsm\ is exactly conformally
invariant on a K3 surface with Ricci-flat metric
\cite{AGG:N=4,Hull:N=4,BS:N=4}. It also shows that the Zamolodchikov
metric and Weyl-Petersson metric coincide exactly on the moduli space
--- a fact which in general $N$=(2,2) theories holds only to leading
order in the large
radius limit \cite{CHS:metric}.

The moduli space of smooth Einstein metrics of volume one on a K3
surface $X$ is determined in \cite{Anderson} to be
$\Gamma_0\backslash\Tch3{19}-\cZ$, where $\Gamma_0=\operatorname{Diff}(X)/
\operatorname{Diff}^0(X)$ is the group of
components of the diffeomorphism group of $X$, and $\cZ$ is the space of
orbifold metrics. (It is
generally believed that such
orbifolds should be included when considering a string target space
\cite{DHVW:}; if we include them, we get the geodesically complete
space $\Gamma_0\backslash\Tch3{19}$.)
The discrete group $\Gamma_0$ is determined in
 \cite{Mat:d,Borcea:d,Don:inv} to coincide with the group
\begin{equation}
O^+(\Lambda^{3,19}):=O(\Lambda^{3,19})\cap O^+(3,19),
\end{equation}
which has index 2 in $O(\Lambda^{3,19})$. The
``missing'' $\Z_2$ in $O(\Lambda^{3,19})$
may be generated by $-I$ acting on $\R^{3,19}$.

We also know that the \nlsm\
on a \CY\ manifold is invariant under all translations $B\to B+v$, where $v\in
H^2(X,\Z)$. Thus we should divide the space $\R^{3,19}$ of $B$-fields
by additive translations by
$\Lambda^{3,19}$.   (As an abstract group, these additive translations
simply form a $\Z^{22}$.)
Furthermore one may consider complex conjugation of the target space.
If one considers such a complex conjugation for a \sm\ one sees that
the transformation $B\to-B$ is required in addition to the conjugation
of the complex structure of the target. Such a transformation may be
represented by $-I$ and thus generates the missing $\Z_2$ from above.
Therefore, in terms of the decomposition (\ref{eq:decomp})
one sees a group $O(\Lambda^{3,19})\ltimes\Lambda^{3,19}$ (which is the full
space-group of $\Lambda^{3,19}$) of identifications
that should be made on the right-hand side and thus on $\Tch4{20}$.
Thus we obtain
\begin{equation}
  \Gamma\supseteq O(\Lambda^{3,19})\ltimes\Lambda^{3,19}. \label{eq:clsym}
\end{equation}

This is the maximal set of identifications that can be made from
classical geometry. Any further statement requires some quantum
geometry. The mirror construction of \cite{GP:orb} provides us with
such a tool. First we need the clarify the meaning of the mirror map
in the context of $N$=(4,4) theories.
Mirror symmetry reverses the sign of a $U(1)$ charge derived from an
$N$=2 chiral algebra.  Now
an $N$=4 chiral algebra
contains an $SU(2)$ affine subalgebra, which is larger than the
$U(1)$ affine subalgebra of an $N$=2 theory.  Given an $N$=4 theory,
though, we may choose an $N$=2
subalgebra of the $N$=4 algebra by specifying the corresponding
$U(1)\subset SU(2)$. Each $N$=4 theory thus gives rise to an
$S^2\cong SU(2)/U(1)$ of $N$=2 theories. This leads to construction of
the moduli space of $N$=(2,2) theories of string on a K3,
which we denote by $\cM_{q\bar q}$, as a fibre bundle
\begin{equation}
  \pi:\;\cM_{q\bar q}\to\cM,
\end{equation}
with fibre $S^2\times S^2$.
Each point in $\cM_{q\bar q}$ corresponds to an $N$=(4,4) theory
on which a particular $N$=(2,2) structure has been chosen;
the subscripts $q$ and $\bar q$ denote
charges with respect to the left and right $U(1)$ currents respectively
within the
$N$=(2,2) theory.

A {\em (left) mirror map}\/ on the corresponding Teichm\"uller
space $\Tch4{20}_{q\bar q}$ of $N$=(2,2) theories is a map
$\mu:\Tch4{20}_{q\bar q}\to\Tch4{20}_{q\bar q}$ with the property
that, for all $S\in\Tch4{20}_{q\bar q}$, the $N$=(2,2) theories
at $S$ and $\mu(S)$ are isomorphic, but with a switch of charge
assignments $(q,\bar
q)\leftrightarrow(-q,\bar q)$. We also define {\em right mirror maps}\/
which switch $(q,\bar q)\leftrightarrow(q,-\bar q)$. Such a map
on $\Tch4{20}_{q\bar q}$  can be pushed down to a map $\bar\mu$ on
the Teichm\"uller space $\Tch4{20}$ ---
since two theories which are mirrors as $N$=(2,2) theories
are isomorphic as $N$=(4,4) theories, any mirror map within $\Tch
4{20}$ should give rise to an element of $\Gamma$. Note that some
mirror maps on $\Tch4{20}_{q\bar q}$ may be trivial on $\Tch4{20}$,
in the sense that they induce the identity element of $\Gamma$.
Such maps relate pairs of points in $\Tch4{20}_{q\bar q}$ which
lie over the same point in the base space $\Tch4{20}$.
What will be of more interest are
the nontrivial mirror maps which map to a nontrivial element of
$\Gamma$. The difference between trivial and nontrivial maps is shown
schematically in figure \ref{fig:f1}.

\iffigs
\begin{figure}
  \centerline{\epsfxsize=14cm\epsfbox{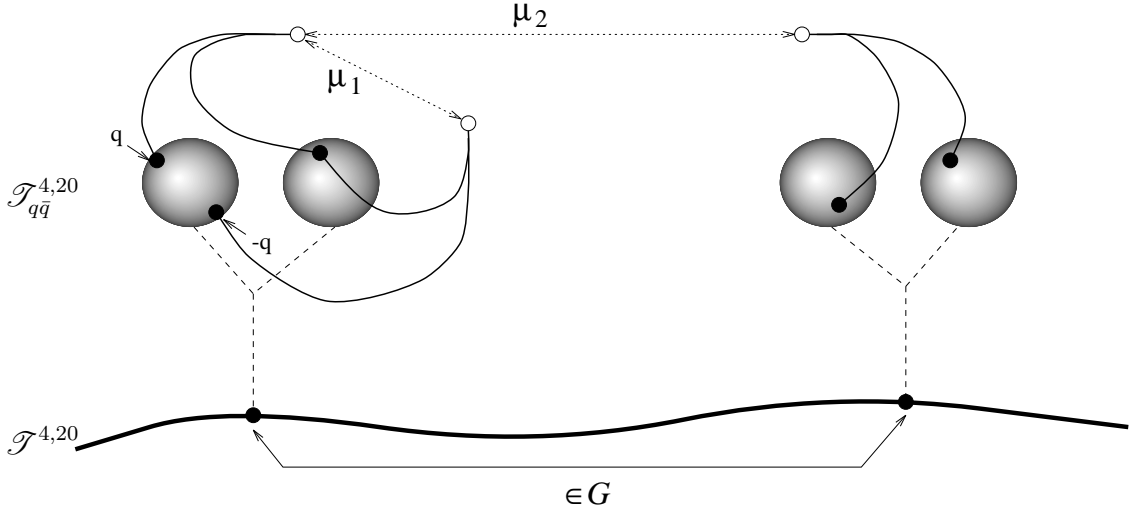}}
\unitlength=1mm \begin{picture}(0,0)
\put(3,45){\makebox{$\Tch4{20}_{q\bar q}$}}
\put(3,12){\makebox{$\Tch4{20}$}}
\end{picture}
  \caption[f1]{Trivial and nontrivial mirror maps. $\Tch4{20}_{q\bar q}$ is
an $(S^2\times S^2)$-bundle over $\Tch4{20}$ as shown. A theory (shown as
an open
circle) is associated with a pair of points, one on each sphere. In
this figure $\mu_1$ is trivial and $\mu_2$ is nontrivial.}
  \label{fig:f1}
\end{figure}
\fi

We now consider an example of a generically nontrivial mirror map with a fixed
point on $\Tch4{20}$. Given the
Gepner model \cite{Gep:} associated via \cite{GVW:} to the K3
hypersurface
\begin{equation}
  X_0^2+X_1^3+X_2^7+X_3^{42}=0 \label{eq:K12Y}
\end{equation}
in the weighted projective space $\P^3_{\{21,14,6,1\}}$ we find the
mirror as an orbifold of the original space by the method of \cite{GP:orb}. The
orbifolding group thus found is trivial and so this theory is its own
mirror.
The fact that this theory is self-mirror is also related to some
issues regarding Arnold's ``strange duality'' observed in \cite{Mart:CCC}.
This mirror map acts non-trivially on the marginal operators of
this theory and thus on the tangent bundle of $\Tch4{20}$.
By analyzing this action, it is possible to show that there is a lattice
$\Lambda^{4,20}\subset\R^{4,20}$ such that the induced automorphism $\bar\mu$
of $\Tch4{20}$ lies in $O(\Lambda^{4,20})$.  In
fact, there
is a decomposition
\begin{equation}
  \Lambda^{4,20}\cong \Lambda^{2,10}\oplus\Lambda^{2,10},
		\label{eq:twobits}
\end{equation}
such that $\bar\mu$ acts by simply exchanging the two terms on
the right-hand side. The decomposition can be chosen so that one of the
$\Lambda^{2,10}$ lattices in (\ref{eq:twobits}) is a sublattice of the
$\Lambda^{3,19}$ lattice on which the classical $O(\Lambda^{3,19})$
symmetry of (\ref{eq:clsym}) acts.

We have thus found the explicit form of another generator of $\Gamma$. A
straight-forward but somewhat involved calculation then shows that the
classical symmetries, $O(\Lambda^{3,19})
\ltimes\Lambda^{3,19}$, and the above $\Z_2$ mirror map
generate the whole of $O(\Lambda^{4,20})$. That is, we have
\begin{equation}
  \Gamma\supseteq O(\Lambda^{4,20}).  \label{eq:Gmax}
\end{equation}
But now from \cite{Allan:}, which establishes that $O(\Lambda^{4,20})$
is a {\em maximal}\/ discrete group acting on $\Tch4{20}$ to yield a
Hausdorff quotient, it follows that
\begin{equation}
  \Gamma\cong O(\Lambda^{4,20}),
\end{equation}
completing the proof that the moduli space has the precise form
\begin{equation}
\cM=O(\Lambda^{4,20})\backslash O(4,20)/\left(O(4)\times O(20)\right)
\end{equation}
(as had been speculated by Seiberg \cite{Sei:K3}).

To summarize, we have shown that the full group of identifications to
be made on the Teichm\"uller space is generated by the following:
\begin{enumerate}
  \item Classical
identifications made for the moduli space of complex structures.
  \item The integral $B$-field shifts.
  \item Complex conjugation (together with a change
in sign of the $B$-field).
  \item The mirror map.
\end{enumerate}
One should note that
this is	 also true for the case $d=1$ where the target space must be a
torus~\cite{DVV:torus}.

It is interesting to note that this moduli space $\cM\cong\Gamma
\backslash\Tch4{20}$ is precisely that which one would obtain for
toroidal compactifications of the heterotic string down to
six-dimensional spacetime using the methods of \cite{N:torus}. This
equivalence between the moduli space of K3 compactifications and
toroidal compactifications (for the relevant strings) was established
locally in \cite{Sei:K3} but it is unclear why they turn out to be
globally isomorphic.


\section{The Space of Total Cohomology}   \label{s:coh}

The derivation of the moduli space $\cM$ in the previous section and
\cite{AM:K3m} is somewhat unpleasant and yet produces a beautifully
symmetric result. This is a consequence of having derived the moduli space
using
the ideas of classical geometry necessitating the intermediate step of
equation (\ref{eq:clsym}). Now that we have the moduli space we may
reinterpret it from the standpoint of quantum geometry in a more
symmetric way.

First let us quickly review the form of the moduli space of complex
structures on a classical K3 surface. Consider a non-vanishing
holomorphic 2-form $\Omega$
on $X$ representing an element of $H^{2,0}(X)$. Let
us specify a ``marking'' of $X$, i.e., a basis $\gamma_i$,
$i=1,\ldots,22$ of $H_2(X,\Z)$ determining an isomorphism of that lattice
with $\Lambda^{3,19}$. The global
Torelli theorem \cite{P-SS:,BR:,Mor:Katata} essentially says
that the complex structure of $X$ is
expressed uniquely in terms of the periods
\begin{equation}
  \int_{\gamma_i}\Omega,
\end{equation}
up to a change of basis of $H_2(X,\Z)$. To state this more
precisely, we write $\Omega=\xi+i\eta$,
where $\xi,\eta\in H^2(X,\R)\cong \Lambda^{3,19}\otimes_{\Z}\R$. The
Hodge-Riemann bilinear relations assert that
\begin{equation}
  \eqalign{
  \langle\Omega,\Omega\rangle&=\int_X\Omega\wedge\Omega=0\cr
  \langle\Omega,\bar\Omega\rangle&=\int_XdV\,\|\Omega\|^2>0.\cr}
\end{equation}
This implies that the oriented 2-plane $\xi\wedge\eta$
in $H^2(X,\R)$ spanned by $\xi$ and
$\eta$ is space-like, i.e., has positive definite metric.
In fact, specifying $\Omega$ up to multiplication by $\C^*$ (which
determines the Hodge structure) is equivalent to specifying the
oriented 2-plane $\xi\wedge\eta$.  Now
the moduli space of oriented space-like 2-planes
in $H^2(X,\R)\cong\R^{3,19}$ is given by
$O(3,19)/\left(SO(2)\times O(1,19)\right)$.
When we mod out by diffeomorphisms, we would expect to obtain a
description of the moduli space of complex structures on a K3 surface
as an open subset of
\begin{equation} \label{eq:cxmod}
  O^+(\Lambda^{3,19})\backslash O(3,19)/\left(SO(2)\times O(1,19)\right)
\end{equation}
(and we would expect to get the entire space (\ref{eq:cxmod}) if we
include orbifold complex structures).  Actually, there are some technical
difficulties in interpreting (\ref{eq:cxmod}) as a moduli space,
but the interpretation is essentially correct (see \cite{Mor:Katata} for
more precise statements).

The extension of this to string theory follows immediately. The
space $\Tch4{20}$ is one connected component of
the set of oriented space-like 4-planes in $\R^{4,20}$.
The space $\R^{4,20}$ can be identified with the space
$H^*(X,\R)$ of all real cohomology groups on $X$ --- the inner product
is obtained by supplementing the intersection pairing on $H^2(X,\R)$
 by a pairing on $H^0(X,\R)\oplus H^4(X,\R)$ which has matrix
(\ref{eq:H}) on the standard generators of those two spaces (i.e., the class
of a point, and the class of the entire space $X$).
The natural lattice $H^*(X,\Z)$ inside of $H^*(X,\R)\cong\R^{4,20}$
inherits an integer-valued pairing from this inner product,
giving it the structure
$\Lambda^{4,20}$.
 An $N$=(4,4) theory
on a K3 surface is thus specified uniquely by an oriented
 space-like 4-plane in
$H^*(X,\R)$ (belonging to the correct component)
where the 4-plane is located relative to the $H^*(X,\Z)$
lattice.

This picture becomes clearer when we go to the $N$=(2,2) moduli space.
Now the moduli space, $\cM_{q\bar q}$, is an $(S^2\times S^2)$-bundle over
$\cM$. In fact one may show that the Teichm\"uller space takes
the form
\begin{equation}
  \Tch4{20}_{q\bar q}=O^+(4,20)/\left(SO(2)\times SO(2)\times O(20)\right).
\end{equation}
In the language of the previous paragraph, this is one component of the space
of
orthogonal pairs of oriented
space-like 2-planes in $\R^{4,20}$. That is, we not only
specify an oriented
space-like 4-plane but we give it an internal structure of being
spanned by two orthogonal oriented
2-planes. We may now specify these oriented 2-planes
in terms of complex vectors as in the classical case. Namely, define
$\Omega$ and $\mho$ as arbitrary elements of $H^*(X,\C)$ (mod $\C^*$)
such that
\begin{equation}
  \eqalign{
  \langle\Omega,\Omega\rangle&=\langle\Omega,\mho\rangle=
  \langle\mho,\mho\rangle=\langle\bar\Omega,\mho\rangle=0,\cr
  \langle\Omega,\bar\Omega\rangle&>0,\quad\langle\mho,\bar\mho\rangle>0.
  \cr} \label{eq:Om}
\end{equation}
The space of such vectors gives precisely the Teichm\"uller space
required. Thus we may identify the moduli space of $N$=(2,2) theories
on a K3 surface as being the space of all possible choices of $\Omega$
and $\mho$ subject to (\ref{eq:Om}) where the positions of these
vectors are located relative to $H^*(X,\Z)$. This statement may be
considered to be the quantum version of the global Torelli theorem.

As explained earlier,  mirror maps will act on the Teichm\"uller space
 $\Tch4{20}_{q\bar
q}$. In fact a nontrivial
mirror map may be identified with the interchange $\Omega
\leftrightarrow\mho$. Our picture now begins to look very similar to
that pursued in \cite{Cand:mir} for the case $d=3$. There is a slight
difference since in the latter case one had
\begin{equation}
  \eqalign{\Omega&\in H^{3,0}\oplus H^{2,1}\oplus H^{1,2}
	\oplus H^{0,3},\cr
   \mho&\in H^{0,0}\oplus H^{1,1}\oplus H^{2,2}
	\oplus H^{3,3}.\cr}
\end{equation}
For the K3 case however, we may only specify the general statement
that both $\Omega$ and $\mho$ lie in the full cohomology
$H^*$ at a generic point in moduli space. The
following construction may be used to reconcile the
approaches. Let us choose two null vectors $v$ and $w$
in $H^*(X,\Z)\cong \Lambda^{4,20}$ such
that $\langle v,w\rangle=1$, so that $v$ and $w$ span a hyperbolic plane.
Let $F$ be
the subspace of $H^*(X,\R)$ perpendicular to $v$ and $w$. For each
oriented space-like 4-plane $\Pi$, we define
\begin{equation}
\eqalign{\Omega&:=\ \mbox{complex vector associated to }\Pi\cap F,\cr
\mho&:=\ \mbox{complex vector associated to }\Omega^\perp\cap\Pi.\cr}
\end{equation}
This
amounts to choosing a specific $N$=(2,2) theory to represent an
$N$=(4,4) theory. We may now identify $F$ with $H^2(X,\R)$. If
$\Omega$ specifies the $H^{2,0}$ direction, (\ref{eq:Om}) then implies
\begin{equation}
  \eqalign{\Omega&\in H^{2,0}\oplus H^{1,1}\oplus H^{0,2},\cr
   \mho&\in H^{0,0}\oplus H^{1,1}\oplus H^{2,2}.\cr}
\end{equation}
The complex structure of $X$ can thus be determined from $\Omega$ in
the usual way. One can also obtain the K\"ahler form and $B$-field
information from $\mho$.  The mirror map interchanges the two.

Note that we may equally specify $\Omega\to\bar\mho,\mho\to\bar\Omega$
as a mirror map. We may compose this with the previous mirror map to
produce the map $\Omega\to\bar\Omega,\mho\to\bar\mho$. This map may be
identified as the map taking $(q,\bar q)\to(-q,-\bar q)$ in the
$N$=(2,2) theory or equivalently as complex conjugation of the target
space. Note that such a map is usually trivial in the moduli space of
$N$=(2,2) theories but that our construction, through specific
labeling of the $U(1)$ charges, leads to a non-trivial map.


\section{Mirror Symmetry for Algebraic K3 Surfaces}      \label{s:alg}

Now that we have identified how mirror maps
act on the moduli space of conformal field theories on a K3 surface,
we can explain several of the mirror-type phenomena which have been
observed for algebraic K3 surfaces.

Given any subgroup $M$ of $H^2(X,\Z)$ such that the cokernel
$H^2(X,\Z)/M$ has no torsion and the intersection form on
$M\otimes\R$ has signature $(1,\rho-1)$, we can associate the
family of ``algebraic K3 surfaces of type $M$'' which consists of
all complex structures on the K3 for which the classes on $M$ are
all of type $(1,1)$.  This construction was introduced into the
mathematics literature \cite{Pinkham} as part of the explanation
of Arnold's ``strange duality'' put forward by Pinkham and Dolgachev,
and moduli spaces of these structures were studied in detail in
\cite{Mor:Tod,MS:}.
The moduli space of algebraic K3 surfaces of type $M$ has complex dimension
$20-\rho$ when $M\otimes\R$ has signature $(1,\rho-1)$.

We can extend this idea, and
define the set of ``CFT's of type $M$'' on a K3 surface
to be the image in the conformal field theory moduli space of the
set of \sm s  whose complex structure is an algebraic K3 surface
of type $M$, and whose K\"ahler class and $B$-field are taken from
the space $M\otimes\R$.
The complexified K\"ahler moduli space
has complex dimension $\rho$, so the entire CFT moduli space of type
$M$ has complex dimension $(20-\rho)+\rho=20$.  Since the construction
segregates complex and K\"ahler deformations, the CFT moduli space of
type $M$ embeds naturally into the $N$=(2,2) moduli space $\cM_{q\bar{q}}$.

Suppose that the orthogonal complement $M^\perp$ can be written in
the form $M^\perp=H\oplus N$ for some lattice $N$ (orthogonal direct
sum).  Then a mirror map $\mu$ can be defined which exchanges the
given hyperbolic plane $H$ with the hyperbolic plane $H^0(X)\oplus H^4(X)$.
When $\mu$ is applied to the CFT moduli space of type $M$, it reverses
the r\^oles of K\"ahler and complex structure deformations.  It is
not hard to see that the resulting moduli space is in fact the CFT
moduli space of type $N$ --- the complex structure is one for which $N$
(rather than $M$) consists of $(1,1)$ classes, and the K\"ahler structure
is now taken from $N\otimes\R$.
This gives a precise CFT interpretation of Arnold's strange duality
and other related phenomena for K3 surfaces (an earlier version of
which was given by Martinec \cite{Mart:CCC}).

Finally we note a link between our construction of the mirror map for
K3 surfaces and Voisin's construction of mirror pairs of \CY\
threefolds \cite{Voi:K3} (see also \cite{Bor:K3}). The latter
construction consists of taking an orbifold $Y=(X\times E)/\Z_2$, where
$E$ is a torus of complex dimension one. It was shown in \cite{Voi:K3}
that for suitable pairs $X_1,X_2$ of $X$, $Y_1$ and $Y_2$ are mirror
pairs at the level of the Hodge numbers, $H^{p,q}(Y_1)=H^{3-p,q}(Y_2)$;
and ``A-model'' and ``B-model'' \cite{W:AB} correlation functions.
The K3 surfaces involved must be of type $M$, with $M$ the lattice
invariant under the $\Z_2$-action.
One may show
that our mirror map for K3 surfaces directly establishes $X_1$ and
$X_2$ as a mirror pair and thus it follows that $Y_1$ and $Y_2$ are
truly mirror manifolds at the level of conformal field theory.


\section*{Acknowledgements}

The work of P.S.A. was supported by  NSF grant PHYS92-45317,
and the work of D.R.M. was supported by an American Mathematical
Society Centennial Fellowship.


\end{document}